\documentclass{article}

\usepackage{graphicx,epsf}
\usepackage{amssymb}

\newcommand{\beq}{\begin{eqnarray}}
\newcommand{\eeq}{\end{eqnarray}}

\newcommand{\btem}{\bibitem}

\title{Lattice Study of Low-lying Nonet Scalar Mesons in Quenched Approximation}

\author{Hiroaki Wada${}^{a}$, Teiji Kunihiro${}^{b}$, Shin Muroya${}^{c}$, \\
 Atsushi Nakamura${}^{d}$, Chiho Nonaka${}^{e}$ \\
and Motoo Sekiguchi${}^{f}$ \\
({\it SCALAR Collaboration}) }
\date{}

\begin{document}

\maketitle

\footnotesize
\begin{center}
${}^{a}$Faculty of Political Science and Economics, Kokushikan University, Tokyo 154-8515, Japan \\
${}^{b}$Yukawa Institute for Theoretical Physics, Kyoto University, Kyoto 606-8502, Japan \\
${}^{c}$Matsumoto University, Matsumoto 390-1295, Japan \\
${}^{d}$RIISE, Hiroshima University, Higashi-Hiroshima 739-8521, Japan \\
${}^{e}$Department of Physics, Nagoya University, Nagoya 464-8602, Japan \\
and \\
${}^{f}$School of Science and Engineering, Kokushikan University, Tokyo 154-8515, Japan
\end{center}

\normalsize

\begin{abstract}
Using lattice QCD simulation in the quenched approximation, 
we study the $\kappa$ meson, which is $^3P_0$ 
in the quark model,  
and compare experimental and other lattice data.  
The $\kappa$ is  the lowest scalar meson with strangeness and 
constitutes the scalar nonet.
The obtained mass is much higher than the recent experimental value, 
and therefore the $\kappa(800)$ is difficult to consider as a simple two-body
constituent-quark structure, 
and may have another unconventional structure.
\end{abstract}

\section{Introduction}

We are in the age of renaissance of hadron
spectroscopy, initiated by the announcement of the pentaquark baryon
\cite{nakano}, which is followed by the discovery of 
many other possible exotic hadrons with a mass larger than 2 GeV 
containing heavy quarks\cite{Quigg:2005tv}.
These experimental developments prompted the intensive theoretical
studies of QCD dynamics with new as well as old ideas on the structure
and dynamics of the exotic hadrons, such as
chiral dynamics\cite{Nowak:1992um},
multi-quark states with diquark correlations or
molecular states and hybrids\cite{Quigg:2005tv}. 

\begin{figure}[htb]
\begin{center}
\includegraphics[width=0.9\linewidth]{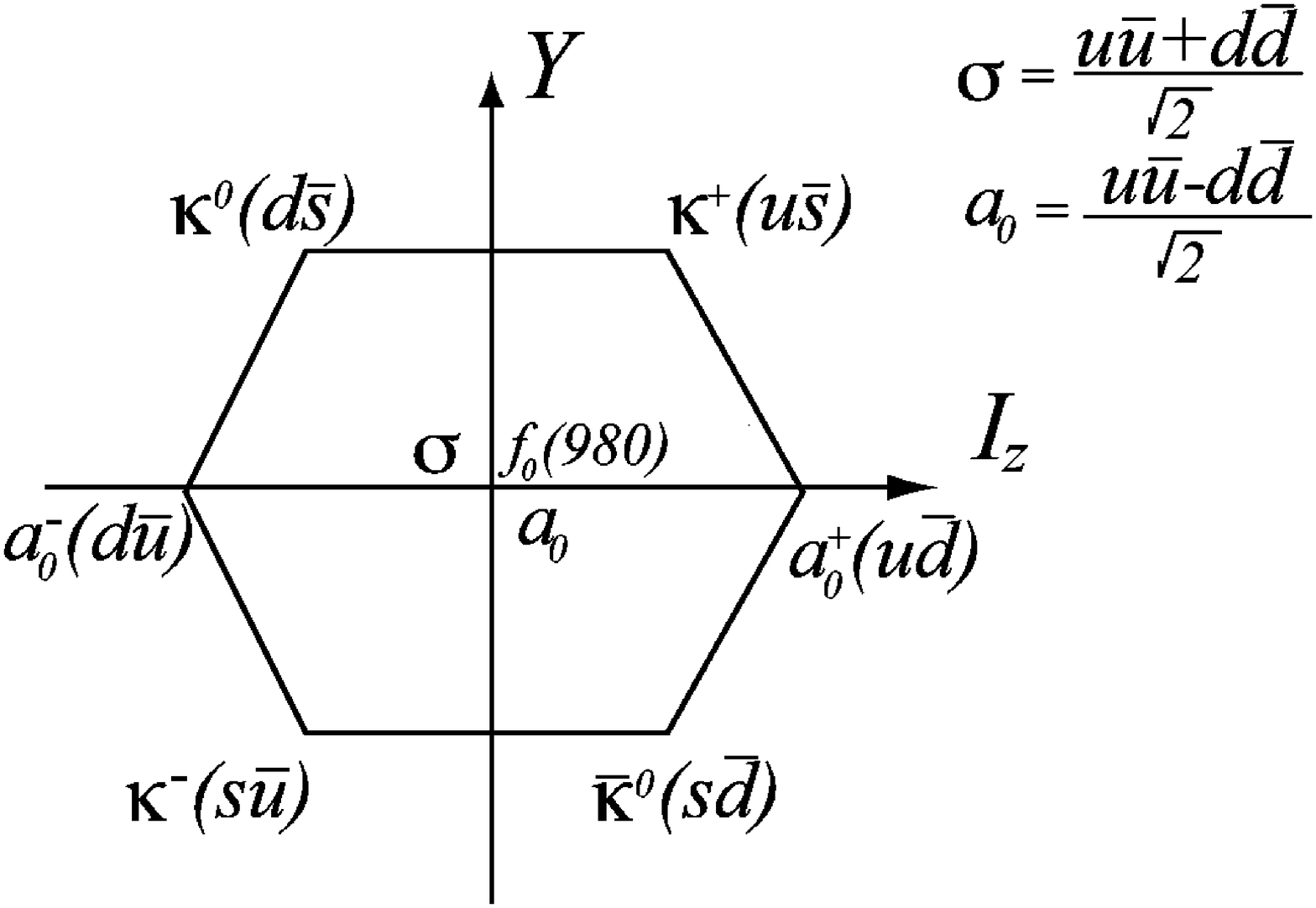}
\caption{
Scalar meson nonet. 
The $\sigma$ and $f_0(980)$ mesons may be ideal mixing
states of singlet state 
$\frac{1}{\sqrt{3}} ( u \bar{u} + d \bar{d} + s \bar{s} )$
and octet state 
$\frac{1}{\sqrt{6}} ( u \bar{u} + d \bar{d} - 2 s \bar{s} )$. 
There is , however, experimental evidence that the $\sigma$ meson
consist of only $u \bar{u}$
and $d \bar{d}$ components. Hence, we take the $\sigma$ wave function given
in the figure.
}
\label{fig:nonet}
\end{center}
\end{figure}

Such a controversy on the structure of hadrons is
also the case  for the scalar mesons below 1 GeV:
the existence of the $I=0$ and $J^{PC}=0^{++}$ meson, i.e.,
the $\sigma(400-600)$, has been reconfirmed \cite{PDG,leutwyler} 
after around twenty years not only in $\pi$$\pi\ $ scattering 
but also in various decay
processes from heavy-quark systems, {\it e.g.} , 
D $\to \pi \pi \pi$ and $\Upsilon(3S) \to \Upsilon \pi \pi$  
\cite{KEK,E791decay,Ishida,Bugg}.
Moreover, the resonance of a scalar meson with $I=1/2$
is also reported to exist in the K-$\pi$ system  
with a mass $m_{\kappa}$ of about 
800 MeV \cite{Bugg,E791,BES}.
This meson is called the $\kappa$ meson and may constitute
the nonet scalar state together with the $\sigma$ meson.
See Fig.\ref{fig:nonet}.

%

The problem is the nature of these low-lying scalar mesons
\cite{close-tornqvist}:
they cannot be ordinary $q\bar{q}$ mesons as described in the 
non-relativistic constituent quark model
since in such a quark model, 
the $J^{PC}$=$0^{++}$ meson is realized in the $^{3}P_{0}$ state, 
which implies that the mass of the $\sigma$ meson must be as high as 1.2
$\sim$ 1.6 GeV.
Thus, the low-lying scalar mesons below 1 GeV 
have been a source of various ideas of exotic structures, as mentioned above:
they may be four-quark states such as $qq\bar{q}\bar{q}$ \cite{Alfold},
or $\pi\pi$ or K$\pi$ molecules
as the recent high-lying exotic hadrons can be.
These mesons may be {\em collective} $q\bar{q}$ states described
as a superposition of  many {\em atomic} $q\bar{q}$ 
states \cite{NJL,HK85}.
A mixing with glueball states is also possible
\cite{Lee,McNeile,Narison,giacosa}. 

In the previous work\cite{ScalarSIGMA1,ScalarSIGMA2,ScalarSIGMA3,ScalarSIGMA4},
we have presented a lattice calculation for the $\sigma$ meson,
by full lattice QCD simulation
on the $8^{3}\times16$ lattice using the plaquette action and
Wilson fermions: We have shown that the disconnected diagram 
plays an essential role in order to make the $\sigma$ meson mass 
light. 
The importance of the disconnected diagram suggests that the
wave function of the $\sigma$ meson may have a significant 
four-quark, a collective $q$-$\bar{q}$
or an even glueball component, although the smallness of the 
lattice requires caution in giving a definite conclusion.
In contrast to the $\sigma$ meson, the $\kappa$ is a flavor non-singlet state
with which a glueball state cannot mix.
In previous reports \cite{ScalarSIGMA4,ScalarKAPPA},
we reported also a preliminary analysis on the $\kappa$ meson 
using the dynamical fermion for 
the $u(d)$ quark but using the valence approximation for the $s$ quark,
which shows that 
the $I=1/2$ scalar meson has a mass as large as about 1.8 GeV and 
cannot be identified with the $\kappa$ meson observed in experiments.

The lattice volume in the previous investigations was admittedly 
too small to yield a definite conclusion at all, and 
 the lattice cutoff was not appropriately chosen to accommodate
large masses : $m_{\kappa}a>1$, where $a$ is the lattice spacing.
Hence, we present a  simulation with weaker couplings
on a larger lattice than any other previous simulations
although in the quenched level.
We perform quenched level simulations on the 
$\kappa$ meson so as {\it to clarify the structure of the mysterious 
scalar meson  rather than to reproduce the experimental
value of the mass}; a precise quenched-level simulation should
give a rather clear perspective on whether
the system can fit with the simple constituent-quark model
picture or not. 

\section{Simulation}

We perform a quenched QCD calculation using the Wilson fermions,
with the plaquette gauge action, on a relatively large lattice
($20^3 \times 24$). 

The values of the hopping parameter for the $u/d$ quark are 
$h_{u/d} = 0.1589, 0.1583$ and 0.1574, while 
$h_s = 0.1566$ and 0.1557 for the $s$ quark.
Using these hopping parameters except for $h_s=0.1557$, 
CP-PACS collaboration performed a quenched QCD calculation of 
the light meson spectrum 
with a larger lattice ($32^3 \times 56$) \cite{CP-PACS}, which we refer to for comparison.
The gauge configurations are generated 
by the heat bath algorithm at $\beta = 5.9$. 
After 20000 thermalization iterations, we start to calculate 
the meson propagators. On  every 2000 configurations,
80 configurations are  used for the ensemble average.

We emply the point-like source and sink for the $\kappa^{+}$ meson
\begin{equation}
\hat{\kappa}(x) \equiv \sum_{c=1}^3\sum_{\alpha=1}^4 
{\bar{s}_\alpha^c(x) u_\alpha^c(x)} \ \ ,
\label{eq:kappa_operator}
\end{equation}
where $u(x)$ and $s(x)$ are the Dirac operators 
for the $u/d$ and $s$ quarks,  and  
the indices $c$ and $\alpha$ 
denote the color and Dirac-spinor indices, respectively.
The point source and sink in Eq.(\ref{eq:kappa_operator}) lead a positive spectral
function $\rho(m^2)$ in the correlation function
 $ \langle \hat{\kappa}(t) \hat{\kappa}(0) \rangle =
\int dm \rho(m^2){\rm exp}(-mt) $.
The result obtained here is thus an upper bound of $\kappa$ mass,
because our result should include excited states.

\begin{figure}[htb]
\begin{center}
\includegraphics[width=\linewidth]{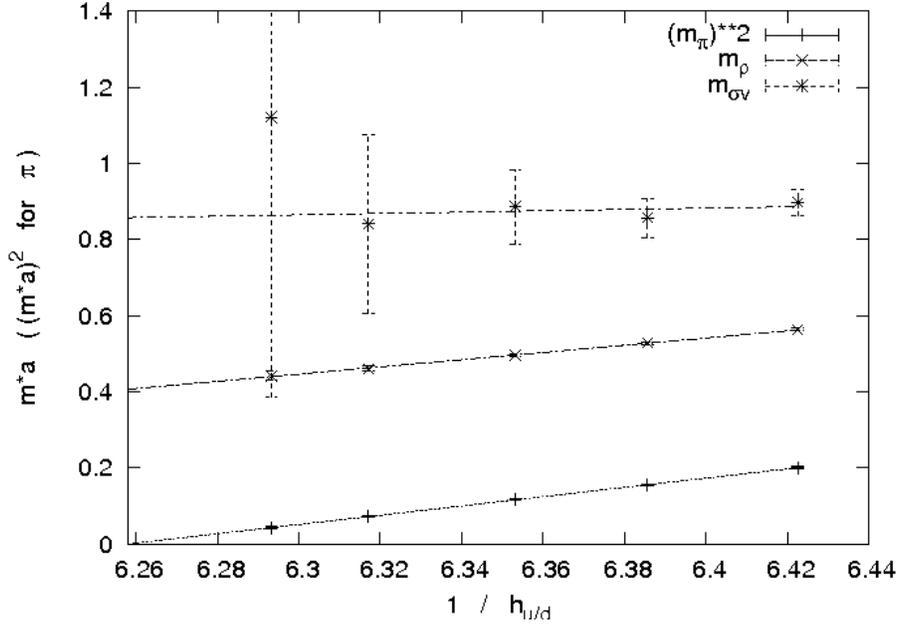}
\caption{$m_{\pi}^2$, $m_{\rho}$ and $m_{\sigma_v}$ in the lattice
unit as a function of the inverse $h_{u/d}$. 
The chiral limit is obtained at $h_{\rm crit}$ = 0.1598(1).}
\label{fig:extrapolation}
\end{center}
\end{figure}

First, we check finite lattice volume effects by comparing
our results for 
the $\pi$ and $\rho$ masses as well as the mass ratio $m_{\pi}/m_{\rho}$ 
with those of the CP-PACS group.  
The results are summarized in Table \ref{table:pi_rho}. 
Our result for the $\rho$ meson mass
is only slightly ($<$ 5 $\%$) larger than the CP-PACS's result.
The resulting larger value is reasonable because
the smaller lattice size gives rise to a mixture of higher mass states.
We rather emphasize that the deviation between our results and the
larger lattice result (CP-PACS) is so small in spite of the large difference 
in the lattice size. 
\begin{center}
\begin{table*}[h]
\caption{Summary of results for $\bar{q}q$ type mesons. }
\label{table:pi_rho}
\begin{tabular}{c|c|c|c|c|c}
\hline
\hline
$h_{u/d}$ &   0.1589   &   0.1583   &   0.1574   &   0.1566   &   0.1557    \\ \hline
$m_{\pi}$          & 0.2064(62) & 0.2691(36) & 0.3401(29) & 0.3935(28) & 0.4478(28)  \\
$m_{\rho}$         &  0.442(13) &  0.461(06) &  0.496(05) &  0.527(04) &  0.563(03) \\
$m_{\pi}/m_{\rho}$ &  0.467(21) &  0.584(10) &  0.686(05) &  0.746(03) &  0.796(03) \\ \hline
$m_{\sigma_v}$ & 1.12(74) & 0.84(23) & 0.886(98) &  0.857(52) & 0.897(35) \\
\hline
\multicolumn{6}{c}{CP-PACS}\cite{CP-PACS}    \\ \hline \hline
$m_{\pi}$ & 0.20827(33) & 0.26411(28) & 0.33114(26) & 0.38255(25) & $-$  \\
$m_{\rho}$ & 0.42391(132) & 0.44514(96) & 0.47862(71) & 0.50900(60) & $-$  \\
$m_{\pi}/m_{\rho}$ & 0.491(2) & 0.593(1) & 0.692(1) & 0.752(1) & $-$    \\ \hline
\end{tabular}
\label{table:qqbar}
\end{table*}
\end{center}
In Fig.~\ref{fig:extrapolation}, 
we show $m_{\pi}^2$, $m_{\rho}$ and $m_{\sigma_v}$ in the lattice unit
as a function of the inverse hopping parameter $1/h_{u/d}$ for the $u/d$ quark. 
The chiral limit ($m_{\pi}^2 = 0$) is obtained
at $h_{u/d}=0.1598(1)\equiv h_{\rm crit}$ ($1/h_{\rm crit}$=6.2581). 
We find the lattice spacing $a$ = 0.1038(33) [fm] in the chiral limit
from the value $m_{\rho}a$ = 0.406(13) at this point 
with the physical $\rho$ meson mass being used for $m_{\rho}$.
Note that these values are consistent with
the  CP-PACS's result, $h_{\rm crit}$ = 0.1598315(68) and $a$ = 0.1020(8) [fm], 
within the error bars. 
%

In Table \ref{table:pi_rho}, the mass of the 
valence $\sigma$ for each
hopping parameter is shown;
the valence $\sigma$, which is denoted as $\sigma_v$,
is defined as the scalar 
meson described solely with the connected propagator. 
The mass ratio $m_{\sigma_v}/m_\rho$ varies from 2.5 ($h_{u/d}=0.1589$) 
to 1.6 ($h_{u/d}=0.1557$), which is consistent with our previous 
results \cite{ScalarSIGMA4}. 
In other words, without the disconnected part of 
the propagator the ``$\sigma$" mass becomes heavy. 

\begin{figure}[htb]
\begin{center}
\includegraphics[width=\linewidth]{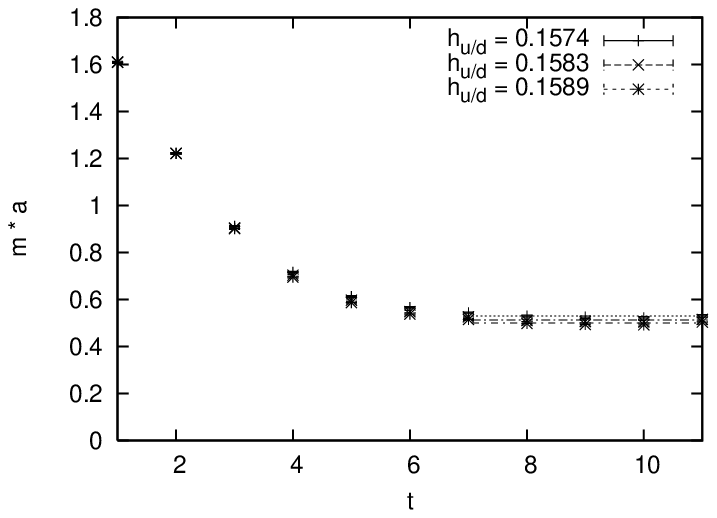}
\caption{Effective mass plots 
of $K^*$ for $s$ quark hopping parameter $h_s$ = 0.1557.}
\label{fig:K*}
\end{center}
\end{figure}

\begin{figure}[htb]
\begin{center}
\includegraphics[width=\linewidth]{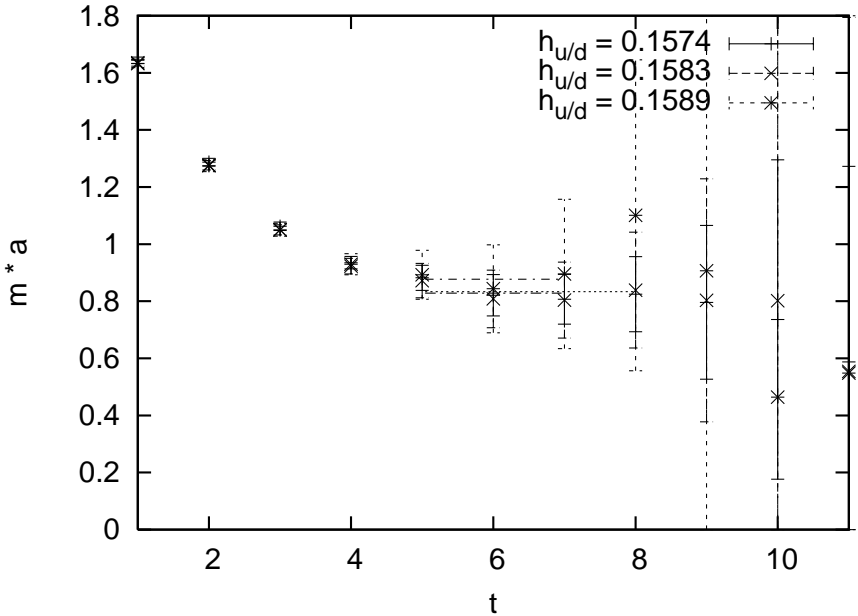}
\caption{Effective mass plots  
of the $\kappa$ meson for the $s$ quark 
hopping parameter $h_s$ = 0.1557.}
\label{fig:kappa}
\end{center}
\end{figure}

\begin{figure}[htb]
\begin{center}
\includegraphics[width=\linewidth]{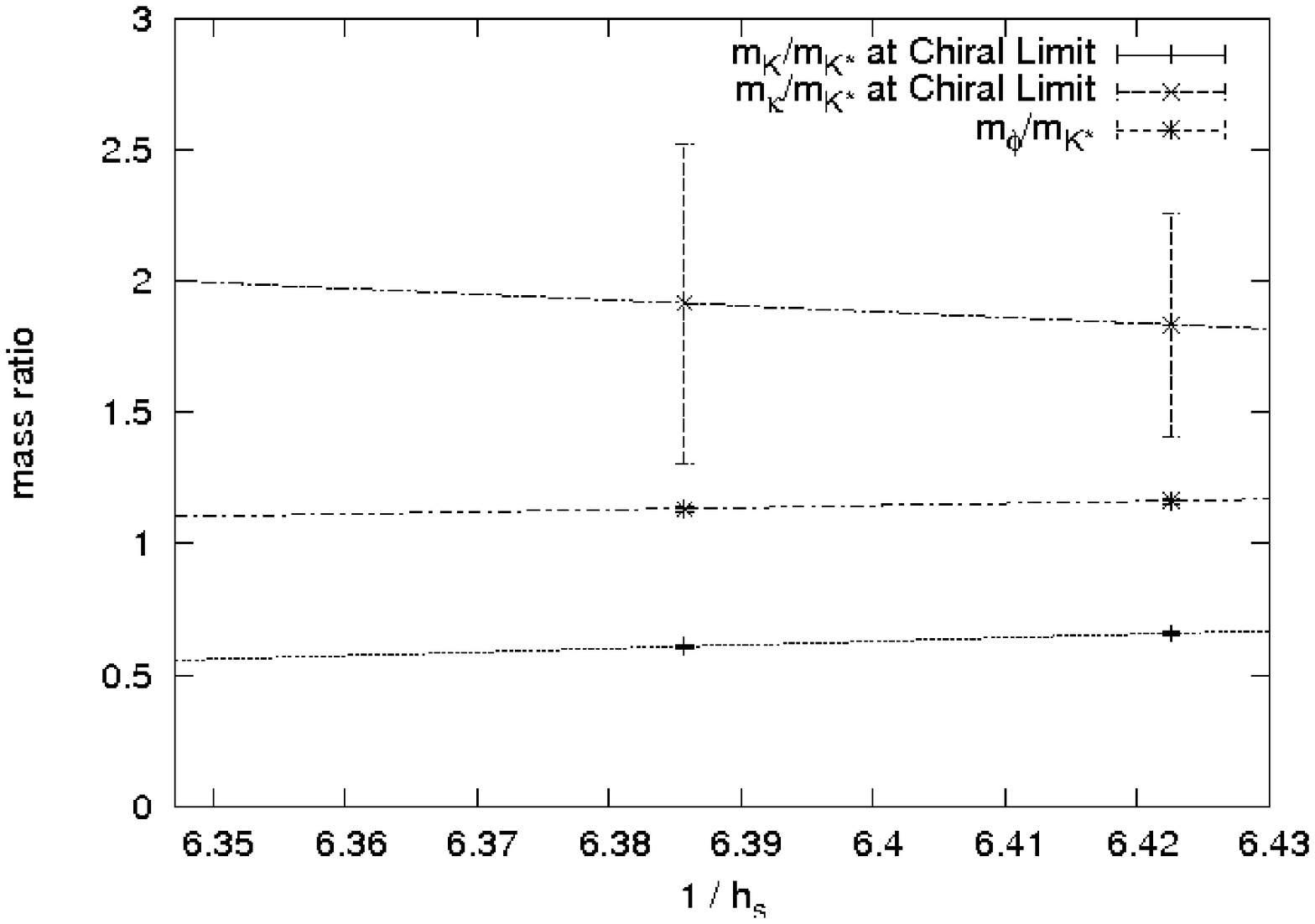}
\caption{The ratios $m_K/m_{K^*}$ and $m_{\kappa}/m_{K^*}$
at chiral limit, and $m_{\phi}/m_{K^*}$ for $s$ quark hopping parameters 
$h_s$ = 0.1566 and 0.1557.}
\label{fig:mass_ratio}
\end{center}
\end{figure}

The propagators of the $K$, $K^*$ and $\kappa$ mesons 
are calculated with the same configurations 
using the $s$-quark hopping parameter, $h_s$ = 0.1566 and 0.1557.
For $h_s$ = 0.1557, the effective mass plots of the $K^*$ and $\kappa$ 
mesons are shown in Figs.~\ref{fig:K*} and \ref{fig:kappa}. 
The masses of the $K$, $K^{*}$ and $\kappa$ mesons, which are 
extracted from the effective mass plots \cite{DeGrand}
, are 
summarized in Tables \ref{table:ratio1566} and 
\ref{table:ratio1557}.
Errors are estimated by jack-knife method.
We find that the effective masses of the $K$ and $K^*$ mesons have 
only small errors and are taken to be reliable, 
while that of the $\kappa$ meson suffers from large errors, especially at
larger time regions.
To avoid possible large errors coming from the data at large $t$,  
we fit the effective mass of the $\kappa$ meson
only in the time range $5 \le t \le 7, 8$ 
where the effective masses are almost constant with small errors.
Since the effective mass of the $K^*$ meson is reliable,
we show the mass of the $\kappa$ in terms of the ratio to  $m_{K^*}$:
Table \ref{table:mass_ratio} gives the mass ratios $m_{K}/m_{K^{*}}$
and $m_{\kappa}/m_{K^{*}}$  at the chiral 
limit together with $m_{\phi}/m_{K^*}$ for $h_s=0.1566$ and 0.1577. 
For example, $m_\kappa/m_{K^*}=0.89(29)/0.4649(69)=1.92(61)$ at 
$h_s=0.1566$ in Table \ref{table:mass_ratio}.  
These calculated mass ratios are shown in Fig.~\ref{fig:mass_ratio}.
All the mass ratios are almost independent of $h_s$.
Although the error bar for $m_\kappa/m_{K^*}$ is 
large, the behavior as a function of $h_s$ is reasonable.        
\begin{table}[htb]
\begin{center}
\caption{Summary of results for the $K$, $K^{*}$ and $\kappa$ mesons at
$h_s$ = 0.1566.}
\label{table:ratio1566}
\begin{tabular}{c|c|c|c|c}
\hline
\hline
$h_{u/d}$   &  $h_{\rm crit}^{1)}$ &   0.1589   &   0.1583   &   0.1574  \\ \hline
$m_{K}$     &   0.2829(23)   &  0.3138(33)  &  0.3368(30)  &  0.3677(29)  \\
$m_{K^{*}}$ &   0.4649(69)   &   0.4821(57)  &  0.4941(49)  &  0.5117(42)  \\
$m_{\kappa}$ &    0.89(29)   &    0.88(23)   &   0.81(12)   &  0.814(81)   \\
\hline
\multicolumn{5}{c}{CP-PACS} \cite{CP-PACS} \\ \hline \hline
$m_{K}$   &   $-$    & 0.30769(28) & 0.32833(26) & $-$  \\
$m_{K^{*}}$ &   $-$    & 0.46724(84) &  0.47749(74)  & $-$   \\ \hline
\end{tabular} \\
\end{center}
$^{1)}$ $h_{\rm crit}$ = 0.1598(1).
\end{table}
\begin{table}[htb]
\begin{center}
\caption{Summary of results for the $K$, $K^{*}$ and $\kappa$ mesons at
$h_s$ = 0.1557.}
\label{table:ratio1557}
\begin{tabular}{c|c|c|c|c}
\hline
\hline
$h_{u/d}$   &  $h_{\rm crit}^{1)}$  &  0.1589   &   0.1583   &   0.1574  \\ \hline
$m_{K}$     &    0.3188(25)   &  0.3474(31) & 0.3684(29) & 0.3971(28) \\
$m_{K^{*}}$ &    0.4835(61)   &  0.5006(52) &  0.5126(44) & 0.5299(37) \\
$m_{\kappa}$ &     0.89(21)  &    0.88(16)  &   0.828(96) &   0.833(72)    \\
 \hline
\end{tabular} \\
\end{center}
$^{1)}$ $h_{\rm crit}$ = 0.1598(1). \\
\end{table}
%
\begin{table}[htb]
\begin{center}
\caption{Summary of results for the mass ratios
$m_K/m_{K^*}$ and $m_\kappa/m_{K^*}$
together with $m_\phi/m_{K^*}$ at chiral limit for $u/d$ quarks.  
}
\label{table:mass_ratio}
\begin{tabular}{c|c|c||c|c}
\hline
\hline
$h_s$                   &  0.1566     & 0.1557    &  0.1563(3)   & 0.1576(2)   \\
$1/h_s$                 &  6.3857     & 6.4226    &  6.396(13)   & 6.3452(80)  \\  \hline
$m_{\phi}/m_{K^*}$      &  1.135(10)  & 1.164(10) & 1.143$^{1)}$ &     $-$ \\
$m_K/m_{K^{*}}$         &  0.6086(79) & 0.6593(63)& 0.623(11)    & 0.5556$^{1)}$  \\
$m_{\kappa}/m_{K^{*}}$  &  1.92(61)   & 1.84(43)  &  1.89(55)     & 2.00(80)  \\
\hline
\end{tabular} \\
\end{center}
$^{1)}$ inputs for calculation of physical value of $h_s$. See the text.  
\end{table}

We have searched for the physical value of the $s$ quark hopping 
parameter $h_s$ in the following two ways, both of 
which are found to give similar results:
1)~ By tracing a regression line 
for $m_{\phi}/m_{K^*}$ (Fig.~\ref{fig:mass_ratio}), we have $h_s$ = 0.1563(3)
(or $1/h_s$ = 6.396(13)) for $m_{\phi}/m_{K*}$=1019[MeV]/892.0[MeV]=1.143 
(input), taken from the PDG \cite{PDG}.
This hopping parameter gives the mass ratio 
$m_{\kappa}/m_{K^*}$ = $1.89(55)$.
2)~ We have also determined the hopping parameter
so as to reproduce the mass ratio 
$m_K/m_{K^*}$ = 495.6[MeV]/892[MeV] = 0.5556, with
$m_K= 495.6$ [MeV] being the average value of the Kaon masses given in the PDG 
\cite{PDG}.
The resulting value is found to be $h_s$ = 0.1576(2) 
(or $1/h_s$ = 6.3452(80)), which in turn gives the mass ratio 
$m_{\kappa}/m_{K^*}$ = 2.00(80).
The mass ratios obtained using methods 1) and 2) are 
also presented in Table \ref{table:mass_ratio}. 
Both methods give almost identical results for the masses 
of the $\kappa$, that are about twice that of the $K^*$.

\section{Concluding remarks}

The motivation of our lattice study is 
to reveal the nature of the scalar meson nonet,
and the results should be important especially 
in clarifying how the $\kappa$ meson with
a reported low mass $\sim 800$ MeV obtained from experiments
can be compatible with the 
valence or constituent quark model:
the $\kappa$ is a
$P$-wave $q\bar{q}$ bound state in the non-relativistic
quark model, and the $\kappa$ meson constitutes
a nonet together with the $\sigma$ meson and the $a_0$ mesons.

There have not been many lattice studies of $\kappa$ meson.
Recently, estimations of the $\kappa$ meson have been reported
by two groups.
Prelovsek {\it et al.} \cite{Sasa2} 
have presented a rough estimate of the mass of
the $\kappa$ as $1.6$ GeV, which is  obtained using the average quark mass
of the $u$ and $s$ quarks from the dynamical simulations
with the degenerate $N_f=2$ quarks on a $16^3\times 32$ lattice.
Mathur {\it et al.} have studied $u\bar{s}$ meson with the overlap fermion in the quenched approximation
 and obtained a mass of the $u\bar{s}$ scalar meson to be 1.41 $\pm$ 0.12 GeV \cite{Mathur}.
The UKQCD Collaboration has
studied to some extent the $\kappa$ meson using the dynamical
$N_{f}$=2 sea  quarks and a valence strange quark
on a $16^3\times 32$ lattice \cite{UKQCDkappa};
they estimated the $\kappa $ mass as about 1.1 GeV,
which is much smaller than those in \cite{ScalarSIGMA4,ScalarKAPPA,Sasa2} 
but still far from the experimental value $\sim$800 MeV.

In this paper, we have presented the lattice simulation results
in the quenched approximation
for the $\kappa$ meson; the results on the $\pi$, $\rho$, $K$
and $K^*$ mesons are also shown for comparison.

We have first checked that 
the  masses of the $\pi$, $\rho$, $K$ and $K^*$ mesons 
obtained in our simulation are in good agreement
with those on a larger lattice ($32^3\times 56$) \cite{CP-PACS}; 
our results are only within five percent larger than the latter.
Our estimated value of the mass of the $\kappa$ is $\sim$ 1.7 GeV, which is 
larger than twice the experimental mass $\sim 800$ MeV. 
This result was expected on the basis of our experience 
in calculating the $\sigma$ meson.
The relatively heavy mass of the $\kappa$ may
be at least partly attributed to the absence of the disconnected diagram in 
the $\kappa$ propagator; the $\kappa$ propagator is composed of only 
a connected diagram.
While the disconnected diagram was
essential for realizing the low-mass $\sigma$ \cite{ScalarSIGMA4}, 
it does not exist for the $\kappa$; therefore, the mass of the $\kappa$
is not made lighter by the disconnected diagram.
Indeed, the mass of the valence 
$\sigma_v$  described solely with the connected propagator 
is far larger than the experimental value
$\sim 500$-$600$ MeV, as seen in Table \ref{table:qqbar}.  

Our lattice study and the quark model analysis\cite{QM85} suggest 
that the simple two-body  constituent-quark picture 
of the $\kappa$ meson does not agree well with 
the experimentally observed $\kappa$.
Note that the quench simulation is a clean theoretical experiment
in which a virtual intermediate like $qq\bar{q}\bar{q}$ is
highly suppressed \cite{Alfold}.
Therefore, 
if its existence with the reported low mass is experimentally established, 
the dynamical quarks may play an essential role for making
the $\kappa$ mass so lighter
or 
the $\kappa$ may contain
an unconventional state such as
a $qq\bar{q}\bar{q}$\cite{tetra} or $K\pi$  molecular
state\cite{torn-phys-rep}, which are missing in the 
calculation here. 

In order to establish this possible scenario, 
the systematic errors should be much reduced 
in future simulations.  
Our statistics here is reasonably high 
(80 configurations separated by 2000 sweeps), 
and the standard meson masses have small error bars; see Fig.\ref{fig:K*}. 
On the contrary, as seen in Fig.\ref{fig:kappa},
the effective mass of  $\kappa$ suffers from large errors 
for large $t$,
which may be due to a small overlap of the physical states.
This is not surprising because $\kappa$ is a P-wave meson, and
expected to be extended.
Choosing more adequate
extrapolation operators and with much higher statistics,
we can study the dynamics of
hadrons by comparing results in 
the quenched lattice QCD,
full lattice QCD and various effective theories/models that include 
the constituent quark models with and without 
the tetra-quark structure, chiral effective theories.

{\bf Acknowledgment}
T.K. is supported by Grants-in-Aid for Scientific Research from 
the Ministry of Education, Culture, Sports, Science and Technology
(No. 17540250) and 
for the 21st century COE ``Center for Diversity and Universality in
Physics'' program of Kyoto University.
The work is partially supported by 
Grants-in-Aid for Scientific Research from
the Ministry of Education, Culture, Sports, Science and Technology
Nos. 13135216 and 17340080.
The calculation was carried out on SX-5 at RCNP, Osaka University and
on SR-8000 at KEK.

\end{document}